\documentclass[preprint,preprintnumbers,amsmath,amssymb,amsfonts,prd,aps,titlepage]{revtex4-1}
\usepackage{epsfig}

\begin{document}

\preprint{UTTG-26-12}
\preprint{TCC-025-12}
\title{Non-Gaussian Correlations Outside the Horizon in Local Thermal Equilibrium}
\author{Joel Meyers}
\email{jmeyers@cita.utoronto.ca}
\affiliation{Canadian Institute for Theoretical Astrophysics, University of Toronto, Toronto, Ontario M5S 3H8, Canada}
\affiliation{Theory Group, Department of Physics, University of Texas, Austin, Texas 78712, USA}
\affiliation{Texas Cosmology Center, University of Texas, Austin, Texas 78712, USA}

\date{\today}

\begin{abstract}
Making a connection between observations of cosmological correlation functions and those calculated from theories of the early universe requires that these quantities are conserved through the periods of the universe which we do not understand.  In this paper, the results of [0810.2831] are extended to show that tree-approximation correlation functions of Heisenberg picture operators for the reduced spatial metric are constant outside the horizon during local thermal equilibrium with no non-zero conserved quantum numbers.
\end{abstract}

\maketitle

\section{Introduction}

If observations of the cosmic microwave background or large scale structure are to tell us anything about the early universe, we need to understand how the correlation functions that we calculate from theories of inflation are related to the correlation functions that are present during the radiation-dominated era at a temperature low enough such that we understand the contents of the universe.  This presents a problem because there are several eras in the early universe which we do not understand including dark matter decoupling and baryon and lepton synthesis.

In \cite{Weinberg:2008si} it was shown that under essentially all conditions the non-linear classical equations governing matter and gravitation in cosmology have adiabatic solutions in which, far outside the horizon, in a suitable gauge, the reduced spatial metric $g_{ij}(\mathbf{x},t)/a^2(t)$ becomes a time-independent function $\mathcal{G}_{ij}(\mathbf{x})$, and all perturbations to the other metric components and to all matter variables vanish.  Furthermore, it was shown that the adiabatic solution is attractive in the case of local thermal equilibrium with no non-zero conserved quantum numbers.  However, this is not enough to conclude the the correlation functions for $\tilde{g}_{ij}(\mathbf{x},t)$ become time-independent at late time.  In calculating non-gaussian correlations of the reduced metric, one generically encounters loop graphs which receive contributions from fields with arbitrarily large wave numbers, invalidating the expansion in powers of $a^{-1}$ as applied to the Heisenberg picture interacting fields \cite{Weinberg:2008nf}.  In fact, this problem arises even for classical fields, where the initial conditions for the non-linear field equations are taken to be stochastic variables.  Furthermore, the prescription for calculating correlation functions even at the leading order in perturbation theory introduces an implicit time-dependence which must be shown to vanish in the appropriate limit if correlation functions truly become constant.  Here we apply the results of \cite{Weinberg:2008mc} to show that in the tree approximation the correlation functions of Heisenberg picture operators for the reduced spatial metric are indeed constant outside the horizon during local thermal equilibrium with no non-zero conserved quantum numbers.

\section{Explicit Adiabatic Solution}
Here we review the results of \cite{Weinberg:2008si} which found a solution of the non-linear classical gravitational field equations for which the reduced metric approaches a constant at late times.  Most of this section simply recalls the arguments given in \cite{Weinberg:2008si}, however we will slightly extend these results by calculating the subleading terms in the momentum potential $U(\mathbf{x},t)$ which we will need in section \ref{correlationssection}.

We will use the ADM parametrization \cite{Arnowitt:1962hi} for the metric:
\begin{align}\label{ADM}
	g_{00}&=-N^2+g_{ij}N^iN^j \, , \qquad g_{0i}=g_{ij}N^j\equiv N_i \, , \nonumber \\
	g^{00}&=-N^{-2} \, , \qquad g^{0i}=N^i/N^2 \, , \qquad g^{ij}={}^{(3)}g^{ij}-N^iN^j/N^2 \, ,
\end{align}
where ${}^{(3)}g^{ij}$ is the reciprocal of the 3 $\times$ 3-matrix $g_{ij}$.  It will be convenient also to write
\begin{equation}\label{reducedmetric}
	g_{ij}(\mathbf{x},t)=a^2(t)\tilde{g}_{ij}(\mathbf{x},t) \, ,
\end{equation}
where $a(t)$ is the Robertson-Walker scale factor appearing in the unperturbed solution.  If we take units with $8\pi G \equiv 1$ the Lagrangian takes the form
\begin{align}
	L &= \frac{1}{2}\int \! d^3x \, \sqrt{-\mathrm{Det} g} \left\{-R^{(4)}\right\} + L_m \nonumber \\
	&=\frac{a^{3}}{2}\int \! d^3x \, N \sqrt{\tilde{g}} \left\{-a^{-2}\tilde{g}^{ij}\tilde{R}_{ij}+C^i_{\phantom{i}j}C^j_{\phantom{j}i}-(C^i_{\phantom{i}i})^2 + 2\mathcal{L}_m \right\} \, , \label{lagrangian1}
\end{align}
and the gravitational field equations are
\begin{equation}\label{momentum}
	\tilde{\nabla}_i\left(C^i_{\phantom{i}j}-\delta^i_{\phantom{i}j} C^k_{\phantom{k}k}\right) = -NT^0_j \, ,
\end{equation}
\begin{equation}\label{energy}
	-a^{-2}\tilde{g}^{ij}\tilde{R}_{ij}-C^i_{\phantom{i}j}C^j_{\phantom{j}i}+(C^i_{\phantom{i}i})^2 = 2N^2T^{00} \, ,
\end{equation}
\begin{align}\label{spatial}
\tilde{R}_{ij}&-C^k_{\phantom{k}k}C_{ij}+2C_{ik}C^k_{\phantom{k}j}+N^{-1}\big(-\dot{C}_{ij}+C^k_{\phantom{k}i}\tilde{\nabla}_jN_k+C^k_{\phantom{k}j}\tilde{\nabla}_iN_k \nonumber\\
&+N^k\tilde{\nabla}_kC_{ij}+\tilde{\nabla}_i\tilde{\nabla}_jN\big)=-T_{ij}+\frac{1}{2}a^2\tilde{g}_{ij}T^{\lambda}_{\phantom{\lambda}\lambda} \, ,
\end{align}
where $\tilde{g}^{ij}(\mathbf{x},t)$ is the reciprocal of the matrix $\tilde{g}_{ij}(\mathbf{x},t)$; $\tilde{R}_{ij}(\mathbf{x},t)$ is the three-dimensional Ricci tensor for the metric $\tilde{g}^{ij}(\mathbf{x},t)$; and $C^i_j(\mathbf{x},t)$ is the extrinsic curvature of the surfaces of fixed time
\begin{equation}\label{extrinsic}
	C^i_{\phantom{i}j}\equiv a^{-2}\tilde{g}^{ik}C_{kj} \, , \qquad C_{ki} \equiv \frac{1}{2N}\left[2a\dot{a}\tilde{g}_{ki}+a^2\dot{\tilde{g}}_{ki}-\tilde{\nabla}_kN_i-\tilde{\nabla}_iN_k\right]\, ,
\end{equation}
where $\tilde{\nabla}_i$ is the three-dimensional covariant derivative calculated with the three-metric $\tilde{g}_{ij}$.

Following \cite{Weinberg:2008si}, we will look for a solution in which $\dot{\tilde{g}}_{ij}$ and $g_{00}+1$ are small perturbations for large $a(t)$, of order $1/a^2(t)$.  We will define space coordinates for which $N^i=0$, so $g_{i0}=0$.  We then have
\begin{equation}\label{extrinsic2}
	C^i_{\phantom{i}j} = H \delta^i_{\phantom{i}j} + \xi^i_{\phantom{i}j}
\end{equation}
where $\xi^i_{\phantom{i}j}$ is, like $\dot{\tilde{g}}_{ij}$ and $\delta N$, a quantity whose leading term is of order $a^{-2}$:
\begin{equation}\label{xi}
	\xi^i_{\phantom{i}j}=\frac{1}{2}\tilde{g}^{ik}\left[\dot{\tilde{g}}_{kj}-2H \delta N \tilde{g}_{kj}\right] + O(a^{-4}) \, .
\end{equation}
The total energy momentum tensor for this solution takes the form
\begin{equation}\label{energymomentum}
	T_{ij}=a^2\bar{p}\tilde{g}_{ij}+\delta T_{ij}\, , \qquad T_{i0}=\delta T_{i0} \, , \qquad T_{00}=\bar{\rho}+\delta T_{00} \, ,
\end{equation}
where the background energy density and pressure are given by $\bar{\rho}=3H^2$, $\bar{p}=-2\dot{H}-3H^2$, and the quantities $\delta T_{ij} /a^2$, $\delta T_{i0}$, and $\delta T_{00}$ are all of order $a^{-2}$.  The gravitational field equations now read
\begin{equation}
	\tilde{\nabla}_i\left(\xi^i_{\phantom{i}j}-\delta^i_{\phantom{i}j}\xi^k_{\phantom{k}k}\right) = \delta T_{0j} + O(a^{-4})\, ,
\end{equation}
\begin{equation}\label{energy2}
	-12H^2\delta N+2\delta T_{00}=-a^{-2}\tilde{g}^{ij}\tilde{R}_{ij}+4H\xi^k_{\phantom{k}k}+O(a^{-4})\, ,
\end{equation}
\begin{align}
\dot{\xi}^i_{\phantom{i}j} &+3H\xi^i_{\phantom{i}j}+H\delta^i_{\phantom{i}j}\xi^k_{\phantom{k}k} +(3H^2-\dot{H})\delta N \delta^i_{\phantom{i}j}=a^{-2}\tilde{g}^{ik}\tilde{R}_{kj} \nonumber \\
&+a^{-2}\tilde{g}^{ik}\delta T_{kj}-\frac{1}{2}a^{-2}\delta^i_{\phantom{i}j}\tilde{g}^{kl}\delta T_{kj}+\frac{1}{2}\delta^i_{\phantom{i}j} \delta T_{00} + O(a^{-4}) \, .
\end{align}

To order $a^{-2}$ adiabatic solutions have no vorticity, so we can introduce a momentum potential $U$, of order $a^{-2}$, such that
\begin{equation}
	\delta T_{i0} = \partial_i U \, .
\end{equation}
The equation of momentum conservation then reads
\begin{equation}\label{momentumconservation}
	a^{-2}\tilde{g}^{ik}\tilde{\nabla}_i\delta T_{kj} = 2\dot{H}\partial_j\delta N+\partial_j(\dot{U}+3HU)+O(a^{-4}) \, .
\end{equation}
We can always write
\begin{equation}
	\delta T_{ij} = a^2(\tilde{g}_{ij}\delta p + \Pi_{ij}) \, ,
\end{equation}
where $\Pi_{ij}$ is a 3-tensor of order $a^{-2}$ representing anisotropic inertia, with $\tilde{g}^{ij}\Pi_{ij}=0$.  For models with vanishing anisotropic inertia, as in the case of local thermal and chemical equilibrium, we have
\begin{equation}\label{bigxi}
\dot{\Xi}^i_{\phantom{i}j}+3H\Xi^i_{\phantom{i}j}=\frac{1}{a^2}\left[\tilde{g}^{ik}\tilde{R}_{kj}-\frac{1}{4}\delta^i_{\phantom{i}j}\tilde{g}^{kl}\tilde{R}_{kl}\right]+O(a^{-4}) \, ,
\end{equation}
where
\begin{equation}\label{bigxidef}
	\Xi^i_{\phantom{i}j} \equiv \xi^i_{\phantom{i}j}+\frac{1}{2}\delta^i_{\phantom{i}j}U \, .
\end{equation}
The general solution of (\ref{bigxi}) is
\begin{align}
	\Xi^i_{\phantom{i}j}(\mathbf{x},t) =& \left[\mathcal{G}^{ik}(\mathbf{x})\mathcal{R}_{kj}(\mathbf{x}) -\frac{1}{4}\delta^i_{\phantom{i}j}\mathcal{G}^{kl}(\mathbf{x})\mathcal{R}_{kl}(\mathbf{x})\right]\frac{1}{a^3(t)}\int_T^t \! a(t') \, dt' \nonumber \\
	&+\frac{B^i_{\phantom{i}j}(\mathbf{x})}{a^3(t)}+O(a^{-4}) \, , \label{bigxisolution}
\end{align}
where $T$ is any fixed time, $\mathcal{G}_{ij}(\mathbf{x})$ is the value of $\tilde{g}_{ij}(\mathbf{x},t)$ at that time, $\mathcal{R}_{ij}(\mathbf{x})$ is the Ricci tensor calculated from the 3-metric $\mathcal{G}_{ij}(\mathbf{x})$; and $B^i_{\phantom{i}j}(\mathbf{x})$ is some function of $\mathbf{x}$ (and $T$), appearing in the solution of the homogeneous equation corresponding to (\ref{bigxi}).

Next, we need to solve for the metric.  From (\ref{xi}), (\ref{energy2}), and (\ref{bigxidef}) we have
\begin{equation}
	\dot{\tilde{g}}_{ij}=2\tilde{g}_{ik}\Xi^k_{\phantom{k}j}+\frac{2H^2}{\dot{H}}\tilde{g}_{ij}\Xi^k_{\phantom{k}k} -\frac{H}{2a^2\dot{H}}\tilde{g}_{ij}\tilde{g}^{kl}\tilde{R}_{kl}+\tilde{g}_{ij}X+O(a^{-4})\, ,
\end{equation}
where $X=O(a^{-2})$ depends on the matter perturbations
\begin{equation}\label{xdef}
	X \equiv -\frac{H}{\dot{H}}\left(\delta T_{00}-2(3H^2+\dot{H})\delta N \right) - U \left(1+\frac{3H^2}{\dot{H}}\right) \, .
\end{equation}
Under a shift $t\rightarrow t+\epsilon(\mathbf{x},t)$ in the time coordinate, with $\epsilon$ of order $a^{-2}$ (and a corresponding transformation $x^i\rightarrow x^i+\mathcal{G}^{ij}\int \! dt \, a^{-2} \partial \epsilon / \partial x^j$ to keep $N_i=0$), the quantity X undergoes the transformation
\begin{equation}
	X \rightarrow X + 2\frac{\partial}{\partial t} \left(\epsilon H\right) + \ldots \, ,
\end{equation}
so we can choose $\epsilon$ to make $X=0$.  This choice does not fix the gauge completely, however.  There is a residual gauge freedom which preserves both $g_{i0}$ and $X=0$ which takes the form
\begin{equation}\label{residual}
	t \rightarrow t + \tau(\mathbf{x})/H(t) \, , \qquad x^i \rightarrow x^i + \mathcal{G}^{ij}(\mathbf{x}) \frac{\partial\tau(\mathbf{x})}{\partial x^i} \int \! \frac{dt}{a^2(t)H(t)} \, ,
\end{equation}
where $\tau$ is an arbitrary function only of $\mathbf{x}$.

With the gauge choice $X=0$ we can solve for $\tilde{g}_{ij}$ without solving for the matter variables to which the metric is coupled.  This solution takes the form
\begin{align}
	\tilde{g}_{ij}(\mathbf{x},t) = &\mathcal{G}_{ij}(\mathbf{x})+2\left[\mathcal{R}_{ij}(\mathbf{x}) -\frac{1}{4}\mathcal{G}_{ij}(\mathbf{x})\mathcal{G}^{kl}(\mathbf{x})\mathcal{R}_{kl}(\mathbf{x})\right] \int_T^t \! \frac{dt'}{a^3(t')} \int_T^{t'} \! a(t'') \, dt'' \nonumber \\
	&+\frac{1}{2}\mathcal{G}_{ij}(\mathbf{x})\mathcal{G}^{kl}(\mathbf{x})\mathcal{R}_{kl}(\mathbf{x}) \int_T^t \! \frac{H^2(t') \, dt'}{\dot{H}(t')a^3(t')} \int_T^{t'} \! a(t'') \, dt'' \nonumber \\
	&-\frac{1}{2}\mathcal{G}_{ij}(\mathbf{x})\mathcal{G}^{kl}(\mathbf{x})\mathcal{R}_{kl}(\mathbf{x}) \int_T^t \! \frac{H(t') \, dt'}{a^2(t')\dot{H}(t')} \nonumber \\
	&+2\mathcal{G}_{ik}(\mathbf{x})B^k_{\phantom{k}j}(\mathbf{x}) \int_T^t \! \frac{dt'}{a^3(t')} +2\mathcal{G}_{ij}(\mathbf{x})B^k_{\phantom{k}k}(\mathbf{x}) \int_T^t \! \frac{H^2(t') \, dt'}{a^3(t')\dot{H}(t')} \nonumber \\
	&+O\left(a^{-4}(t)\right) \, , \label{metricsolution}
\end{align}
where $T$ is again any fixed time, and $\mathcal{G}_{ij}(\mathbf{x})$ and $\mathcal{R}_{ij}(\mathbf{x})$ are the values of $\tilde{g}_{ij}$ and the associated Ricci tensor at that time.

Next we need to solve for the remaining metric component $g_{00}=-N^2$.  From the perfect-fluid form of the energy momentum tensor, we can write
\begin{equation}\label{zerozero}
	\delta T_{00} = \delta \rho +6H^2 \delta N + O(a^{-4})
\end{equation}
so that (\ref{energy2}) and (\ref{bigxidef}) give
\begin{equation}\label{deltarho}
	2\delta \rho=-a^{-2}\tilde{g}^{ij}\tilde{R}_{ij}+4H\Xi^k_{\phantom{k}k}-6HU+O(a^{-4}) \, ,
\end{equation}
while (\ref{zerozero}) and (\ref{xdef}) with the gauge condition $X=0$ give
\begin{equation}\label{deltarhodeltan}
	0=-\frac{H}{\dot{H}}\left(\delta \rho -2\dot{H}\delta N\right)-U\left(1+\frac{3H^2}{\dot{H}}\right)+O(a^{-4}) \,  .
\end{equation}
With vanishing anisotropic inertia (\ref{momentumconservation}) gives
\begin{equation}\label{momentumconservation2}
	0=-\delta p +2\dot{H}\delta N+3HU+\dot{U}+O(a^{-4}) \, .
\end{equation}
We now have three relations for the four quantities $\delta N$, $\delta \rho$, $\delta p$, and $U$, so we need one additional relation to calculate all four.

In local thermal equilibrium with no non-zero conserved quantum numbers the pressure and energy density in any gauge are functions only of the temperature, so
\begin{equation}\label{thermalequilibrium}
	\delta p = \left(\frac{\dot{\bar{p}}}{\dot{\bar{\rho}}}\right)\delta \rho = \left(-1-\frac{\ddot{H}}{3H\dot{H}}\right)\delta \rho \, .
\end{equation}
Combining (\ref{deltarho})
-(\ref{thermalequilibrium}) we arrive at a differential equation for the momentum potential $U$
\begin{align}\label{udifferential}
	\dot{U}+U\left(\frac{\dot{H}}{H}-\frac{\ddot{H}}{\dot{H}}\right) =& -\left(2+\frac{\ddot{H}}{3H\dot{H}}\right) \left(-\frac{1}{2a^2}\tilde{g}^{ij}\tilde{R}_{ij}+2H\Xi^i_{\phantom{i}i}\right) \nonumber \\
	=&-\left(1+\frac{\ddot{H}}{6H\dot{H}}\right)\left(-\frac{1}{a^2}+\frac{H}{a^3} \int_T^t \! a(t') \, dt' \right) \mathcal{G}^{ij}\mathcal{R}_{ij} \nonumber \\
	&-\left(4H+\frac{2\ddot{H}}{3\dot{H}}\right)\frac{1}{a^3}B^i_{\phantom{i}i}+O(a^{-4}) \, ,
\end{align}
where T may be taken as any time during the period of thermal equilibrium, most conveniently at its beginning, and $\mathcal{G}_{ij}$ and $\mathcal{R}_{ij}$ are the reduced metric $g_{ij}/a^2$ and the associated Ricci tensor at that time.  The solution of this equation is
\begin{align}
	U(\mathbf{x},t) =& f(\mathbf{x})\frac{\dot{H}(t)}{H(t)} \nonumber \\
	&-\mathcal{G}^{ij}(\mathbf{x})\mathcal{R}_{ij} (\mathbf{x})\frac{\dot{H}(t)}{H(t)} \int_T^t \! dt' \, \left(\frac{H(t')}{\dot{H}(t')}\right) \left(1+\frac{\ddot{H}(t')}{6H(t')\dot{H}(t')}\right) \nonumber \\
	&\times \left(-\frac{1}{a^2(t')}+\frac{H(t')}{a^3(t')} \int_T^{t'} \! a(t'') \, dt'' \right) \nonumber \\
	&-B^i_{\phantom{i}i}(\mathbf{x})\frac{\dot{H}(t)}{H(t)} \int_T^t \! \frac{dt'}{a^3(t')} \, \left(\frac{H(t')}{\dot{H}(t')}\right) \left(4H(t')+\frac{2\ddot{H}(t')}{3\dot{H}(t')}\right) \nonumber \\
	&+ O(a^{-4}) \, , \label{usolution}
\end{align}
where $f(\mathbf{x})$ is an arbitrary function of position.  We can slightly simplify this expression by using the relation
\begin{equation}\label{totalderivative}
	\frac{d}{dt'}\left(\frac{H(t')}{a^3(t')\dot{H}(t')}\right) = \frac{1}{a^3(t')}-\frac{3H^2(t')}{a^3\dot{H}(t')}-\frac{H(t')\ddot{H}(t')}{a^3(t')\dot{H}^2(t')} \, ,
\end{equation}
so the fourth line of (\ref{usolution}) becomes
\begin{align}
	&-\frac{2H(T)}{3a^3(T)\dot{H}(T)}B^i_{\phantom{i}i}(\mathbf{x}) \frac{\dot{H}(t)}{H(t)} + \frac{2}{3a^3(t)}B^i_{\phantom{i}i}(\mathbf{x}) \nonumber \\
	&- \frac{2}{3}B^i_{\phantom{i}i}(\mathbf{x})\frac{\dot{H}(t)}{H(t)} \int_T^t \! \frac{dt'}{a^3(t')} \, \left(1+3\frac{H^2(t')}{\dot{H}(t')}\right) \, .
\end{align}
The first term can be absorbed into a redefinition of the arbitrary function $f(\mathbf{x})$.  The leading term in (\ref{usolution}), representing a solution of the homogeneous equation corresponding to (\ref{udifferential}), is of zeroth order in $a^{-1}$, and so does not become small for large $a$.  However, this term can be removed by the residual gauge transformation (\ref{residual}) under which
\begin{equation}\label{utransformation}
	U(\mathbf{x},t)\rightarrow U(\mathbf{x},t)+\left(\frac{2\dot{H}(t)}{H(t)}\right)\tau(\mathbf{x}) \, .
\end{equation}
By choosing $\tau(\mathbf{x})$ to have the value $-f(\mathbf{x})/2$, we can cancel the first term in (\ref{usolution}).  After this gauge transformation, $U$ takes the form
\begin{align}
	U(\mathbf{x},t)=& -\mathcal{G}^{ij}(\mathbf{x})\mathcal{R}_{ij} (\mathbf{x})\frac{\dot{H}(t)}{H(t)} \int_T^t \! dt' \, \left(\frac{H(t')}{\dot{H}(t')}\right) \left(1+\frac{\ddot{H}(t')}{6H(t')\dot{H}(t')}\right) \nonumber \\
	&\times \left(-\frac{1}{a^2(t')}+\frac{H(t')}{a^3(t')} \int_T^{t'} \! a(t'') \, dt'' \right) \nonumber \\
	&+\frac{2}{3a^3(t)}B^i_{\phantom{i}i}(\mathbf{x}) - \frac{2}{3}B^i_{\phantom{i}i}(\mathbf{x})\frac{\dot{H}(t)}{H(t)} \int_T^t \! \frac{dt'}{a^3(t')} \, \left(1+3\frac{H^2(t')}{\dot{H}(t')}\right) \nonumber \\
	&+ O(a^{-4}) \, , \label{usolution2}
\end{align}
The leading term of the remainder is of order $a^{-2}$ and thus vanishes for large $a$.

The remaining perturbations $\delta g_{00}=-2\delta N$ and $\delta \rho$ are algebraically related to $U$ by (\ref{deltarho}) and (\ref{deltarhodeltan}), which give
\begin{align}\label{deltarhosolution}
	\delta \rho(\mathbf{x},t)=&-3H(t)U(\mathbf{x},t) \nonumber \\
	&+\frac{1}{2}\mathcal{G}^{ij}(\mathbf{x})\mathcal{R}_{ij}(\mathbf{x}) \left(-\frac{1}{a^2(t)}+\frac{H(t)}{a^3(t)} \int_T^t \! dt' \, a(t') \right)+O(a^{-4})
\end{align}
and
\begin{align}\label{deltansolution}
	2\dot{H}(t)\delta N(\mathbf{x},t)=&\left(\frac{\dot{H}(t)}{H(t)}\right)U(\mathbf{x},t)  \nonumber \\
	&+\frac{1}{2}\mathcal{G}^{ij}(\mathbf{x})\mathcal{R}_{ij}(\mathbf{x}) \left(-\frac{1}{a^2(t)}+\frac{H(t)}{a^3(t)} \int_T^t \! dt' \, a(t') \right)+O(a^{-4}) \, .
\end{align}
Both are of order $a^{-2}$, so for local thermal equilibrium with no non-zero conserved quantum numbers, the adiabatic solution of the non-linear classical field equations is attractive for large $a$.

\section{Tree-Approximation Correlation Functions}\label{correlationssection}
The results of \cite{Weinberg:2008si}, which we reviewed in the previous section, show that the non-linear classical field equations have an adiabatic solution in which the reduced spatial metric is constant outside the horizon and all perturbations to the other metric components and the matter variables vanish, and this solution is attractive during local thermal equilibrium with no non-zero conserved quantum numbers.  However, this is not enough to conclude that correlation functions of the reduced spatial metric are conserved outside the horizon.  In calculating correlation functions of Heisenberg picture interacting fields, one generally encounters loop graphs which receive contributions from fluctuations with arbitrarily small wave number, invalidating the expansion in powers of $a^{-1}$.  In this section we will apply the general theorem of \cite{Weinberg:2008mc} to show that in the tree approximation, correlation functions of the reduced spatial metric are constant outside the horizon during local thermal equilibrium with no non-zero conserved quantum numbers.  This section will closely follow the analysis of \cite{Weinberg:2008nf} which showed that tree approximation correlations are constant during single field inflation.

The generating function for correlation functions of $\tilde{g}_{ij}$ at a time $t_1$ is
\begin{equation}\label{generatingfunction}
	\mathrm{exp}\left\{W[J,t_1]\right\}\equiv\left\langle 0,\mathrm{in}\left|\mathrm{exp}\left[\int \! d^3x \, \tilde{g}^H_{ij}(\mathbf{x},t_1) J^{ij}(\mathbf{x})\right]\right|0,\mathrm{in}\right\rangle \, ,
\end{equation}
where $\tilde{g}^H_{ij}(\mathbf{x},t)$ is the Heisenberg-picture quantum mechanical operator corresponding to $\tilde{g}_{ij}(\mathbf{x},t)$.  Correlation functions for $\tilde{g}_{ij}(\mathbf{x},t)$ are calculated by the formula
\begin{equation}\label{correlations}
	\left\langle 0,\mathrm{in} \left| \tilde{g}^H_{ij}(\mathbf{x},t_1)\tilde{g}^H_{kl}(\mathbf{y},t_1)\ldots \right| 0, \mathrm{in} \right\rangle = \left[\frac{\partial^n}{\partial J^{ij}(\mathbf{x})\partial J^{kl}(\mathbf{x})\ldots} \,  \mathrm{exp}\left\{W[J,t_1]\right\}\right]_{J=0} \, .
\end{equation}

In order to calculate the generating function $W$ in the tree approximation, we need to construct complex \textit{c-number} metric fields $\tilde{g}_{ij}(\mathbf{x},t)$ together with a complex auxiliary field $N(\mathbf{x},t)$, satisfying the following constraints:

\noindent \textbf{(A)} The fields satisfy the Euler-Lagrange equations.  For this case, they are (\ref{momentum})-(\ref{spatial}).

\noindent \textbf{(B)} The fields $\tilde{g}_{ij}$ satisfy constraints at the time $t_1$ when the correlation functions are evaluated
\begin{align}
	&\mathrm{Im} \, \tilde{g}_{ij}(\mathbf{x},t_1)=0 \, , \label{constraint1} \\
	&\mathrm{Im}\left\{\frac{\delta L\left[\tilde{g},\dot{\tilde{g}},t_1\right]}{\delta\dot{\tilde{g}}_{ij}(\mathbf{x},t_1)}\right\} = -J^{ij}(\mathbf{x}) \, . \label{constraint2}
\end{align}

\noindent \textbf{(C)} $\tilde{g}_{ij}$ satisfies a positive frequency constraint at time $t\rightarrow -\infty$, so that it behaves as a superposition of terms proportional to $\mathrm{exp}(-i\omega t)$, with $\omega$ various \textit{positive} frequencies. (This final constraint would be modified if one considered a non-trivial initial state for the cosmological fluctuations.)

With $\tilde{g}_{ij}(\mathbf{x},t)$ and $N(\mathbf{x},t)$ calculated subject to these constraints, the contribution of connected tree graphs to the generating function is given by
\begin{equation}\label{treegeneratingfunction}
	W[J,t_1]_{\mathrm{tree}}=\int_{-\infty}^{t_1} \! \mathrm{Im} \, L\left[\tilde{g}(t),\dot{\tilde{g}}(t),t\right] \, dt + \int \! d^3x \, J^{ij}(\mathbf{x})\tilde{g}_{ij}(\mathbf{x},t_1) \, .
\end{equation}
These constraints give the functions $\tilde{g}_{ij}$ an implicit dependence on the time $t_1$ when the correlations are evaluated, so in order to conclude that the correlation functions are truly time-independent, we must show that the constraints become independent of $t_1$ for sufficiently large $a(t_1)$.  Furthermore, we must show that the time integral appearing in (\ref{treegeneratingfunction}) converges for sufficiently large $a(t_1)$.

For large $a(t_1)$, the constraint (\ref{constraint1}) provides the $t_1$-independent condition that leading term $\mathcal{G}_{ij}(\mathbf{x})$ in the solution for the reduced spatial metric (\ref{metricsolution}) must be real for all $\mathbf{x}$.  The associated Ricci tensor $\mathcal{R}_{ij}(\mathbf{x})$ is then also real, so all of the terms of order $a^{-2}$ appearing in (\ref{metricsolution}) are real.  As a result
\begin{align}
	\mathrm{Im} \, \tilde{g}_{ij}(\mathbf{x},t) =& 2\mathcal{G}_{ik}(\mathbf{x}) \, \mathrm{Im} \, B^k_{\phantom{k}j}(\mathbf{x}) \int_T^t \! \frac{dt'}{a^3(t')} 
	+ 2\mathcal{G}_{ij}(\mathbf{x}) \, \mathrm{Im} \, B^k_{\phantom{k}k}(\mathbf{x})\int_T^t \frac{H^2(t') \, dt'}{a^3(t')\dot{H}(t')} \nonumber \\ &+ O\left(a^{-4}(t)\right) \, , \label{imaginarymetric}
\end{align}
and the leading terms in $\mathrm{Im} \, \tilde{g} (\mathbf{x},t_1)$ are of order $a^{-3}$.

We will next check the convergence of the time integral appearing in (\ref{treegeneratingfunction}) for large $t_1$, returning below to the functional derivative appearing in the constraint (\ref{constraint2}).  In order to do this, we need to say something about the matter Lagrangian $L_m$ appearing in (\ref{lagrangian1}).  As long as comoving entropy density is conserved, which is guaranteed by the condition of local thermal equilibrium with negligible chemical potentials, apart from some total derivative terms the value of matter Lagrangian is given simply by \cite{Schutz:1970my,Schutz:1977df,Brown:1992kc}
\begin{equation}\label{matterlagrangian}
	L_m = \int \! d^3x \, \sqrt{-\mathrm{Det} g} \, p = a^3 \int \! d^3x \, N \sqrt{\tilde{g}} \, \left[\bar{p}+\delta p\right] \, ,
\end{equation}
where $p$ is the pressure.

By examining (\ref{metricsolution}), (\ref{usolution2}), (\ref{thermalequilibrium}), (\ref{deltarhosolution}), and (\ref{deltansolution}) we can see that each of $\dot{\tilde{g}}_{ij}$, $U$, $\delta p$, and $\delta N$ are quantities of order $a^{-2}$ with leading imaginary part of order $a^{-3}$.  Any second order function of these quantities will thus have leading imaginary part $a^{-2} \times a^{-3}$.  

We will first consider the terms in the Lagrangian (\ref{lagrangian1}) containing either 0 or 1 space or time derivative and which are of zeroth order in $\delta N$ and $\delta p$.  These terms are
\begin{align}
	L_0\left[\tilde{g},\dot{\tilde{g}},t\right] &= \frac{a^3}{2} \int \! d^3x \, \sqrt{\tilde{g}} \, \left[-6H^2-2H\tilde{g}^{ij}\dot{\tilde{g}}_{ij} + 2\bar{p} \right] \nonumber \\
	&= \frac{a^3}{2} \int \! d^3x \, \sqrt{\tilde{g}} \, \left[-12H^2-4\dot{H}-2H\tilde{g}^{ij}\dot{\tilde{g}}_{ij} \right] \, ,
\end{align}
where we have used the fact $\bar{p}=-2\dot{H}-3H^2$.   As was pointed out in \cite{Maldacena:2002vr}, this is a total time derivative
\begin{equation}\label{l0}
	L_0\left[\tilde{g},\dot{\tilde{g}},t\right] = -2 \frac{d}{dt} \left( a^3 H \int \! d^3x \, \sqrt{\tilde{g}} \right) \, .
\end{equation}
It was shown in the appendix of \cite{Weinberg:2008nf} that a total time derivative in the Lagrangian has no effect on the correlation functions, so we can ignore these terms in what follows.  Next, we will consider the terms of zeroth order in $\dot{\tilde{g}}_{ij}$ and which are of first order in $\delta N$ or $\delta p$.  These terms are
\begin{align}
	L_1\left[\tilde{g},\dot{\tilde{g}},t\right] &= \frac{a^3}{2} \int \! d^3x \, \sqrt{\tilde{g}} \, \left[6H^2\delta N + 2\bar{p} \delta N + 2\delta p \right] \nonumber \\
	&= \frac{a^3}{2} \int \! d^3x \, \sqrt{\tilde{g}} \, \left[-4\dot{H} \delta N + 2\delta p \right] \, .
\end{align}
We can add to the Lagrangian a term
\begin{align}\label{deltal1}
	\Delta L_1\left[\tilde{g},\dot{\tilde{g}},t\right] &= - a^3  \int \! d^3x \, \sqrt{\tilde{g}} \, \left(3HU+\dot{U}+\frac{1}{2}U\tilde{g}^{ij}\dot{\tilde{g}}_{ij}\right) \nonumber \\
	&= -\frac{d}{dt} \left( a^3  \int \! d^3x \, \sqrt{\tilde{g}} \, U\right) \, ,
\end{align}
which is a total derivative so it does not alter the correlation functions.  Combining this with $L_1$, we have
\begin{align}
	(L_1+\Delta L_1)\left[\tilde{g},\dot{\tilde{g}},t\right] &= \frac{a^3}{2} \int \! d^3x \, \sqrt{\tilde{g}} \, \left[-4\dot{H} \delta N + 2\delta p - 6HU-2\dot{U}-U\tilde{g}^{ij}\dot{\tilde{g}}_{ij} \right] \nonumber \\ 
	&= -\frac{a^3}{2} \int \! d^3x \, \sqrt{\tilde{g}} \, \left(U\tilde{g}^{ij}\dot{\tilde{g}}_{ij}+O(a^{-4})\right) \, ,
\end{align}
where we have used (\ref{momentumconservation2}) to simplify the second line, and the $O(a^{-4})$ term is a second order function of $U$, $\delta N$, and $\delta p$.  The leading imaginary part of $L_1+\Delta L_1$ is thus of order $a^{-2}$.  Next, we will consider the term in $L$ containing the spatial curvature.  By examining (\ref{metricsolution}) and (\ref{deltansolution}), we can see that the leading imaginary part of the term
\begin{equation}
	L_\mathrm{s}\left[\tilde{g},\dot{\tilde{g}},t\right] = -\frac{a^3}{2} \int \! d^3x \, N \sqrt{\tilde{g}} \, a^{-2} \tilde{g}^{ij}\tilde{R}_{ij}
\end{equation}
is of order $a^{-2}$.  All of the remaining terms in the Lagrangian are given by $a^3$ times some second order function of $\dot{\tilde{g}}_{ij}$, $U$, $\delta p$, and $\delta N$, plus terms of higher order in $a^{-1}$.  As a result, all of the remaining terms have a leading imaginary part of order $a^{-2}$.  We therefore conclude that the time integral appearing in (\ref{treegeneratingfunction}) converges to a finite limit for large $t_1$ as $\int^{t_1} \! a^{-2}(t) \, dt$.

Finally, we will consider the functional derivative appearing in the constraint (\ref{constraint2}).  For this purpose, we will add to the Lagrangian the total derivatives $-L_0$ and $\Delta L_1$ identified above in (\ref{l0}) and (\ref{deltal1}), so we have
\begin{align}\label{lagrangianplusderivatives}
	&\frac{\delta (L-L_0+\Delta L_1)\left[\tilde{g}(t),\dot{\tilde{g}}(t),t\right]}{\delta \dot{\tilde{g}}_{ij}(\mathbf{x},t)} \nonumber \\
	&\qquad=\frac{a^3(t)\sqrt{\tilde{g}(\mathbf{x},t)}}{2}\tilde{g}^{ik}(\mathbf{x},t)\left(-2H(t)\delta^j_{\phantom{j}k} + \xi^j_{\phantom{j}k}(\mathbf{x},t) - \delta^j_{\phantom{j}k}\xi^l_{\phantom{l}l}(\mathbf{x},t)\right) \nonumber \\
	 & \qquad \, -\frac{\delta L_0\left[\tilde{g}(t),\dot{\tilde{g}}(t),t\right]}{\delta \dot{\tilde{g}}_{ij}(\mathbf{x},t)} + \frac{\delta \Delta L_1\left[\tilde{g}(t),\dot{\tilde{g}}(t),t\right]}{\delta \dot{\tilde{g}}_{ij}(\mathbf{x},t)} \\ 
	 & \qquad  = \frac{a^3(t)\sqrt{\tilde{g}(\mathbf{x},t)}}{2}\tilde{g}^{ik}(\mathbf{x},t)\left(-U(\mathbf{x},t)\delta^j_{\phantom{j}k} + \xi^j_{\phantom{j}k}(\mathbf{x},t) - \delta^j_{\phantom{j}k}\xi^l_{\phantom{l}l}(\mathbf{x},t)+O(a^{-4})\right) \, . \nonumber
\end{align}
Recalling the definition (\ref{bigxidef}), we can rewrite this as
\begin{align}
	&\frac{\delta (L-L_0+\Delta L_1)\left[\tilde{g}(t),\dot{\tilde{g}}(t),t\right]}{\delta \dot{\tilde{g}}_{ij}(\mathbf{x},t)} \nonumber \\
	&\qquad = \frac{a^3(t)\sqrt{\tilde{g}(\mathbf{x},t)}}{2}\tilde{g}^{ik}(\mathbf{x},t)\Big[\Xi^j_{\phantom{j}k}(\mathbf{x},t) - \delta^j_{\phantom{j}k}\Xi^l_{\phantom{l}l}(\mathbf{x},t)+O(a^{-4})\Big] \, .
\end{align}
The terms of order $a^{-2}$ appearing in (\ref{bigxisolution}) are real, so the leading imaginary part of $\Xi^i_{\phantom{i}j}$ is given by
\begin{equation}\label{imaginarybigxi}
	\mathrm{Im} \, \Xi^i_{\phantom{i}j}(\mathbf{x},t) = a^{-3}(t) \, \mathrm{Im} \, B^i_{\phantom{i}j}(\mathbf{x}) + O\left(a^{-4}(t)\right) \, .
\end{equation}
As a result, we have
\begin{align}\label{constraint2result}
	&\mathrm{Im} \, \frac{\delta (L-L_0+\Delta L_1)\left[\tilde{g}(t),\dot{\tilde{g}}(t),t\right]}{\delta \dot{\tilde{g}}_{ij}(\mathbf{x},t)} \nonumber \\
	&\qquad = \frac{\sqrt{\mathcal{G}(\mathbf{x})}}{2}\mathcal{G}^{ik}(\mathbf{x}) \, \mathrm{Im} \, \left(B^j_{\phantom{j}k}(\mathbf{x})-\delta^j_{\phantom{j}k} B^l_{\phantom{l}l}(\mathbf{x})\right) + O\left(a^{-1}(t)\right) \, ,
\end{align}
and so the constraint (\ref{constraint2}) becomes independent of $t_1$ for large $a(t_1)$.
	
We have shown that during local thermal equilibrium with no non-zero conserved quantum numbers the tree-approximation generating function $W_{\mathrm{tree}}[J,t_1]$ converges to a $t_1$-independent function for large $a(t_1)$, and so the correlation functions calculated in the tree approximation are constant far outside the horizon.



\section{Conclusions}
We have used the results of \cite{Weinberg:2008si} and \cite{Weinberg:2008mc} to show that the tree-approximation correlation functions of the reduced spatial metric $\tilde{g}_{ij}$ are constant outside the horizon during local thermal equilibrium with no non-zero conserved quantum numbers.  This result shows that if we are able to calculate correlation functions from a particular theory of inflation and follow the evolution up until a phase of local thermal equilibrium, then we may compare the predictions of the theory with observations made today, even without an understanding of several phases of the early universe which follow inflation, as long as loop corrections can be neglected.  It has been shown that loop corrections are much smaller than the leading contributions in typical theories of inflation \cite{Weinberg:2005vy,Weinberg:2006ac}, and so the tree approximation is sufficient for making a connection with observations.	

The results presented here are exactly what would be expected from the intuition that the evolution in the tree approximation should correspond to the classical solution.  On the other hand, the explicit proof given here guarantees the constancy of cosmological correlation functions without reference to the classicality of the relevant quantities.

\section*{Acknowledgements}
I would like to thank Eiichiro Komatsu and Steven Weinberg for helpful discussions.  This material is based upon work supported by the National Science Foundation under Grant Number PHY-0969020, and by the Texas Cosmology Center, which is supported by the College of Natural Sciences and the Department of Astronomy at the University of Texas at Austin and the McDonald Observatory.

\bibliography{correlation}

\end{document}